\documentclass[fleqn,twoside]{article}

\usepackage{color}

\usepackage{espcrc2}
\usepackage{graphicx}

\readRCS
$Id: espcrc2.tex,v 1.2 2004/02/24 11:22:11 spepping Exp $
\usepackage{epsfig}

\newcommand{\AmS}{{\protect\the\textfont2
  A\kern-.1667em\lower.5ex\hbox{M}\kern-.125emS}}

\title{
\vspace*{-18mm}
\rightline{
{\normalsize{IPPP/06/59}}}
\vspace*{-2mm}
\rightline{\normalsize{August 2006}}
\vspace*{+6mm}
Third-order QCD results on form factors and coefficient functions%
{\thanks{Presented by A.V. at {\it{Loops and Legs in Quantum Field 
Theory}}, 23--28 April 2006, Eisenach (Germany).}}}

\author{A. Vogt\address{IPPP, Physics Department, Durham University,
 South Road, Durham DH1 3LE, United Kingdom}, 
 S. Moch\address[DESYZ]{DESY, Platanenallee 6, D--15735 Zeuthen, 
 Germany}
 and J.A.M. Vermaseren\address{NIKHEF, Kruislaan 409, 1098 SJ Amsterdam, 
 The Netherlands}} 
       
\def\z#1{{\zeta_{#1}}}

\def\c3a{{C^{\, 3}_{\!A}}}

\def\nf{{n^{}_{\! f}}}
\def\n2f{{n^{\,2}_{\! f}}}
\def\dabc2{{d^{\:\!abc}d_{abc}}}

\newcommand{\hspn}{{\hspace{-1mm}}}

\newcommand{\lsim}{\raisebox{-0.07cm}{$\:\stackrel{<}{{\scriptstyle
 \sim}}\: $} }
\newcommand{\beq}{\begin{equation}}
\newcommand{\eeq}{\end{equation}}
\newcommand{\bea}{\begin{eqnarray}}
\newcommand{\eea}{\end{eqnarray}}
\newcommand{\nn}{\nonumber}

\newcommand{\as}{\alpha_{\rm s}}

\newcommand{\ra}{\rightarrow}
\newcommand{\DD}{{\cal D}}
\newcommand{\ep}{\varepsilon}

\begin{document}

\begin{abstract}
We summarize recent higher-order QCD results based, directly or indirectly, on 
the Mellin-space computation of the next-to-next-to-leading-order splitting 
functions governing the evolution of hadronic parton distributions.  
Specifically, we briefly present third-order results for the coefficient 
functions in inclusive deep-inelastic scattering (including the structure
function $F_3$ not published so far), the on-shell quark and gluon form 
factors, and the total cross section for Higgs production at hadron colliders.
\end{abstract}

\maketitle

\section{Introduction} 

For the next decade, the highest-energy experiments in particle physics will be 
performed at the \mbox{(anti-)}proton--proton colliders {\sc Tevatron} and LHC.
The cross section for, for example, the inclusive production of a high-$p_T^{}$
hadron $h$ at such machines can be schematically written as
\beq
\label{eq:pp-fact}
 \sigma_{ pp\,\ra\, h+X} \: =\: \sum_{f,f'\!,f'\!'} f_p \ast f'_p \ast 
 \hat{\sigma}^{\,\rm f'\!'}_{\rm f\:\!f'} \ast D_{\rm f'\!'}^{\, h} \:\: . 
\eeq
Here $f_p$ stands for the universal momentum distributions of the partons $f$ 
in the proton $p$, $f = q_i,\bar{q}_i, g$ with $i = 1,\ldots ,\nf$, where $\nf$ 
is the number of effectively massless flavours, and $D_{\rm f'\!'}^{\, h}$ are 
the corresponding fragmentation functions of the final-state hadron $h$.
$\hat{\sigma}$ represents the partonic cross section for the process under 
consideration, and power corrections have been disregarded.
 
\vspace{0.5mm}
Hence all quantitative collider studies of the standard model, and of expected 
and unexpected new particles, require a precise understanding of the partonic 
luminosities and of the QCD corrections to the hard cross~sections. For many 
important processes, like Higgs-boson production, at least the second-order 
(NNLO) QCD corrections need to be included, i.e., the third term in 
\beq
\label{eq:sig-exp}
 \hat{\sigma}\: = \: a_{\rm s}^{\:\!n}\,
   \left[ \, \hat{\sigma}^{(0)} + a^{}_{\rm s}\, \hat{\sigma}^{(1)}
   + \, a_{\rm s}^2\, \hat{\sigma}^{(2)} + \ldots \,\right] \:\: .
\eeq
The consistent inclusion of $\hat{\sigma}^{(2)}$ requires parton distributions, 
and in the special case of Eq.~(\ref{eq:pp-fact}) also fragmentation functions, 
evolved with the corresponding NNLO splitting functions
\beq
\label{eq:P-exp}
 P_{\:\!\rm f\:\!f'}^{\,\rm NNLO} \: =\:  
     a^{}_{\rm s} P_{\:\!\rm f\:\!f'}^{(0)} 
   + a_{\rm s}^2  P_{\:\!\rm f\:\!f'}^{(1)} 
   + a_{\rm s}^3  P_{\:\!\rm f\:\!f'}^{(2)} \:\: .
\eeq
 
\vspace{0.5mm}
The computation of the third-order splitting functions for the (unpolarized)
parton distributions has been completed two years ago 
\cite{Moch:2004pa,Vogt:2004mw}, see also Refs.~\cite{Vogt:2004gi,Moch:2004sf}.
As briefly recalled below, that calculation was set up such that it could be
extended to the third-order coefficient functions in inclusive deep-inelastic
scattering (DIS) \cite{Moch:2004xu,Vermaseren:2005qc}. Here we briefly discuss
these coefficient functions, as well as some further results based on their 
calculation \cite {Moch:2005id,Moch:2005tm,Moch:2005ky}, see also
Refs.~\cite{Moch:2005ba,Moch:2005dx,Vogt:2005dw}. A discussion of related NNLO 
results~\cite{Mitov:2006ic} for the evolution of fragmentation functions can 
be found in Ref.~\cite{Mitov:2006zz}.

\section{Third-order DIS coefficient functions}

A calculation of inclusive DIS, $\{\gamma,W\}^{\ast}(q)+ f(p) \!\ra\! X $ with 
$ Q^2 \equiv - q^2 > 0$ and \mbox{$p^2 = 0$}, has been performed in Refs.~\cite 
{Moch:2004pa,Vogt:2004mw,Vermaseren:2005qc} up to the third order in the strong
coupling $a_{\rm s} = \as /(4\pi)$. 
The results have been obtained first for all even or odd values of the Mellin
variable $N$, via the optical theorem and the three-loop forward Compton 
amplitudes $\{\gamma,W\}^{\ast}(q) + f(p) \ra \{\gamma,W\}^{\ast}(q) + f(p)$. 
From these expression the complete Bjorken-$x$ space results have been obtained
by an automated Mellin inversion \cite{Remiddi:1999ew,Moch:1999eb}.

\vspace{0.5mm}
This approach had the crucial advantage of allowing to check the
extensive and involved FORM \cite{Vermaseren:2000nd,Vermaseren:2002rp} codes, 
at almost any stage, by falling back to the {\sc Mincer} program 
\cite{Gorishnii:1989gt,Larin:1991fz} employed in the previous fixed-$N$ 
calculations of Refs.~\cite{Larin:1994vu,Larin:1997wd,Retey:2000nq}. An 
independent check of the $\gamma^\ast$-exchange non-singlet results for 
$N\! =\! 16$ has been performed in Ref.~\cite{Blumlein:2004xt}.

\vspace{0.5mm}
The pole terms of the unfactorized coefficient functions, supplemented by a 
corresponding calculation of DIS by exchange of a scalar $\phi$ directly 
coupling to gluons (like the Higgs boson in the heavy top-quark limit), deliver
the complete set of NNLO splitting functions \cite{Moch:2004pa,Vogt:2004mw}.
The number of relevant diagrams is shown in Table 1. 
The finite pieces of the three-loop $\gamma^{\ast}\!f$ amplitudes, when 
calculated also for the $p^{\,\mu} p^{\,\nu}$ projection of the hadronic 
tensor, lead to the $\as^3$ coefficient functions for $F_L$ and $F_2$ in 
electromagnetic DIS \cite{Moch:2004xu,Vermaseren:2005qc}. 

\vspace{0.5mm}
The former coefficient functions complete the NNLO description of photon-%
exchange DIS in massless perturbative QCD, see, e.g., Ref.~\cite{Martin:2006qv}.
At large Bjorken-$x$ the latter quantities dominate the N$^3$LO corrections for
$F_2$ \cite{vanNeerven:2001pe,Baikov:2006zz}, thus facilitating improved 
determinations of $\as$ from data on deep-inelastic 
scattering. For recent fit-analyses see Ref.~\cite{Blumlein:2006be} and 
references therein. 

\begin{table}[bht]
\vspace*{-5mm}
\[ \begin{tabular}{l c c c c c}
\hline\\ & & & & & \\[-7mm] 
{} &{tree} &1-loop &2-loop &3-loop& proj. \\[1mm]
\hline\\ & & & & & \\[-7mm]
q$\gamma$ & 1  & 3  & \phantom025         & \phantom0\phantom0359 & 2 \\
g$\gamma$ & {} & 2  & \phantom017         & \phantom0\phantom0345 & 2 \\
h$\gamma$ & {} & {} & \phantom0\phantom02 & \phantom0\phantom0\phantom056 & 2 \\
q$W$      & 1  & 3  & \phantom032         & \phantom0\phantom0589 & 1 \\
q$\phi$   & {} & 1  & \phantom023         & \phantom0\phantom0696 & 1\\
g$\phi$   & 1  & 8  & 218                 & \phantom06378 & 1 \\
h$\phi$   & {} & 1  & \phantom033         & \phantom01184 & 1 \\[1mm]
\hline\\ & & & & & \\[-6.5mm]
sum       & 4  & 23 & 394                 & 10367 \\[1mm]
\hline\\
\end{tabular} \]
\vspace*{-3mm}
\caption{Numbers of diagrams employed in the calculation of the three-loop 
 splitting functions and the corresponding coefficient functions.
 $h$ denotes the standard ghost used to simplify the gluon polarization sum.
 The overall sums include the number of Lorentz projections (last column)
 required for extracting the coefficient functions for $F_2$ and $F_L$.}
\end{table}

\begin{figure}[htbp]
\vspace{-8mm}
\centerline{\epsfig{file=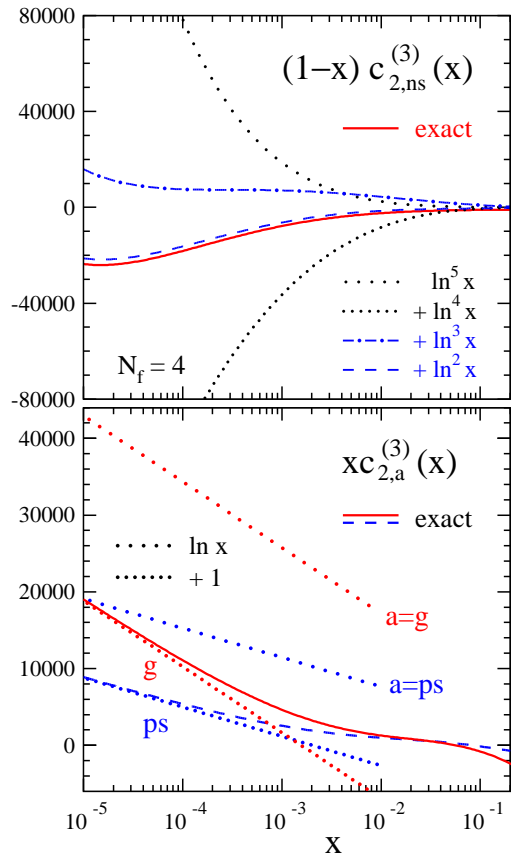,width=6.8cm,angle=0}\quad}
\vspace{-8mm}
\caption{\label{mmvv-f1}
 The small-$x$ behaviour of the third-order flavour non-singlet (top) and
 singlet (bottom) coefficient functions for the structure function $F_2$ in 
 electromagnetic DIS. Also shown are the respective approximations obtained
 by successively including the terms leading for $x \ra  0$.}
\end{figure}

\begin{figure}[htbp]
\vspace{-6mm}
\centerline{\epsfig{file=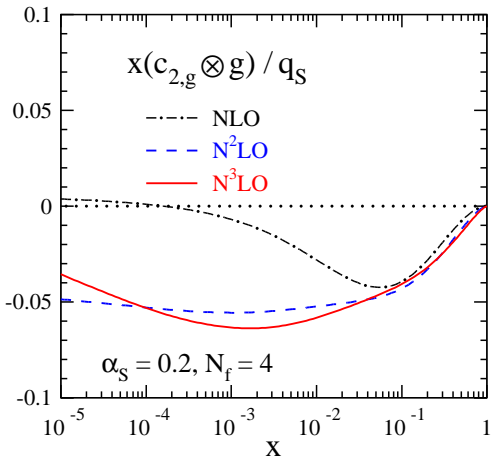,width=6.8cm,angle=0}\quad}
\vspace{-9mm}
\caption{\label{mmvv-f2}
 The perturbative expansion up to three loops (N$^3$LO) of the gluon 
 contribution to the flavour-singlet structure function $F_2$ for $\as = 0.2$,
 $\nf = 4$ and $ xg = 1.6\, x^{\, -0.3} (1-x)^{4.5}\, 
 (1 - 0.6\, x^{\, 0.3\,}) $. The results have been normalized to the 
 leading-order result given by the singlet quark distribution
 $ xq_{\rm s} = 0.6\, x^{\, -0.3} (1-x)^{3.5}\, (1 + 5.0\, x^{\, 0.8\,}) $.
 }
\vspace{-4mm}
\end{figure}

The small-$x$ behaviour of the coefficient functions for $F_2$ is illustrated
in Figs.~\ref{mmvv-f1} and \ref{mmvv-f2}. The former shows the new N$^3$LO 
contributions $c_{\,2}^{(3)}(x)$, as in Eq.~(\ref{eq:sig-exp}) using the 
expansion parameter $a_{\rm s} = \as /(4\pi)$, together with successive 
approximations by the dominant contributions for $ x\ra 0$. Even disregarding 
the additional effect of the ubiquitous Mellin convolution, cf., e.g., 
Ref.~\cite{Vogt:2004mw}, the leading logarithms do not provide a useful 
approximation in the $x$-range accessible to colliders. In particular, a 20\% 
accuracy for $c_{\,2,\rm ns}^{(3)}(x)$ is reached with one and two small-$x$ 
logarithms only at $x < 10^{-50}$ (sic) and $x < 10^{-14}$, respectively.

The perturbative stability at small $x$ is illustrated in Fig.~\ref{mmvv-f2} 
for the gluonic contribution at $Q^2 \approx 30 \ldots 50 \mbox{ GeV}^2$. 
The third-order correction is far smaller than the second-order contribution 
first calculated in Ref.~\cite{Zijlstra:1991qc}, exceeding 1\% of the 
leading-order result only at $\,x \! < \! 2\cdot 10^{-5}\,$. 
It should be noted, on the other hand, that the perturbative expansion appears
to become unstable at $x \lsim 10^{-4}$ at very low scales, $Q^2 \approx 2 
\mbox{ GeV}^2$ \cite{Moch:2004xu,Vermaseren:2005qc}, due to the larger $\as$ 
and especially the flatter small-$x$ shape of the parton distributions.  

\begin{figure}[thb]
\vspace{-1mm}
\centerline{\epsfig{file=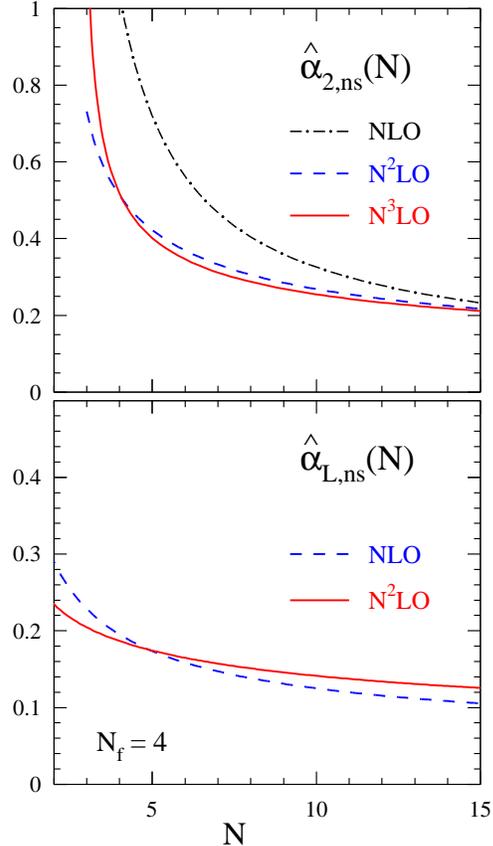,width=6.8cm,angle=0}\quad}
\vspace{-9mm}
\caption{\label{mmvv-f3}
 The $N$-dependent values of $\as$ at which the effect of the $n$-th order
 (N$^n$LO for $F_2$, N$^{n-1}$LO for $F_L$) non-singlet coefficient functions
 is half as large as that of the previous order.}
\vspace{-6mm}
\end{figure}

The convergence of the perturbation series at large-$x\,/\,$large-$N$ 
can be conveniently illustrated by displaying the value $\widehat{\alpha}_{a}
^{\,(n)}(N)$ for which the effect of the $n$-th order term $c_{\:\!a}^{(n)}(N)$ 
is half as large as that of $c_{\:\!a}^{(n-1)}(N)$. This quantity would be 
order-independent for a geometric series, while a systematic decrease with 
increasing $n$ would indicate the asymptotic character of the expansion. 
$\widehat{\alpha}_{a,\rm ns}^{\,(n)}(N)$ are shown in Fig.~\ref{mmvv-f3} for
$a=2,\,L$ (note the different scales). No sign is observed of an imminent 
breakdown of the perturbative expansion at phenomenologically relevant values 
of $N$. 

The third-order coefficient function has also been computed for the (odd-$N$) 
non-singlet quantity $F_3^{\:\!\nu+\bar{\nu}}$, the $qW$ entry in Table 1.
The lengthy exact results and a full discussion will be presented elsewhere 
\cite{Vermaseren:2006zz}. Here we only provide a compact $x$-space 
parametrization, which is sufficiently accurate for all numerical purposes and 
which can be easily transformed to Mellin space for use which complex-$N$ codes
like Ref.~\cite{Vogt:2004ns}. This parametrization is given in 
Eq.~(\ref{eq:c33par}) using the abbreviations $x_1 =  1-x$, $L_0 = \ln\, x$, 
$L_1 = \ln\, x_1$, and $\DD_k = [ x_1^{-1} L_1^k ]_+$ for the usual 
+-distributions. The factor $f\:\!\!l_{02}$ ($=1$ for the numerical evaluation) 
indicates the $\dabc2$ contribution entering at this order for the first time,
cf.~Ref.~\cite{Retey:2000nq}.
\begin{table*}[thb]
\vspace{-2mm}
\begin{eqnarray}
\label{eq:c33par}
  c_{\,3,\,\nu+\bar{\nu}}^{\,(3)}(x)\!\! & \cong &  \quad
       512/27\: \DD_5 - 5440/27\: \DD_4 + 501.099\: \DD_3
     + 1171.54\: \DD_2 - 7328.45\: \DD_1 + 4442.76\: \DD_0
  \nn \\ && \mbox{} \quad
     - 9172.68\: \delta (x_1)
     - 512/27\: L_1^5 + 8896/27\: L_1^4 - 1396\: L_1^3
     + 3990\: L_1^2 + 14363\: L_1
  \nn \\ & & \mbox{} \quad
     - 1853 - 5709\: x + x\,x_1 (5600 - 1432\: x)
     - L_0 L_1 (4007 + 1312\: L_0) - 0.463\: xL_0^6
  \nn \\ & & \mbox{} \quad
     - 293.3\: L_0
     - 1488\: L_0^2 - 496.95\: L_0^3 - 4036/81\: L_0^4 - 536/405\: L_0^5
  \nn \\[1mm] &+& \mbox{} \nf \:\big\{ \,
     640/81\: \DD_4 - 6592/81\: \DD_3 + 220.573\: \DD_2
     + 294.906\: \DD_1 - 729.359\: \DD_0 + 2575.46\: \delta (x_1)
  \nn \\ & & \mbox{} \quad
     - 640/81\: L_1^4 + 32576/243\: L_1^3 - 660.7\: L_1^2 + 959.1\: L_1
     + 516.1 + x\,x_1 (635.3 + 310.4\: x)
  \nn \\ & & \mbox{} \quad
     - 465.2\: x + 31.95\: x_1 L_1^4
     + L_0 L_1\, (1496 + 270.1\:L_0 - 1191\: L_1)
     - 1.200\: xL_0^4 + 366.9\: L_0
  \nn \\ & & \mbox{} \quad
     + 305.32\: L_0^2 + 48512/729\: L_0^3 + 304/81\: L_0^4 \, \big\}
  \nn \\[1mm] &+& \mbox{} \nf^{\!\!\! 2} \:\big\{ \,
     64/81\: \DD_3 - 464/81\: \DD_2 + 7.67505\: \DD_1 + 1.00830\: \DD_0
     - 103.2602\: \delta (x_1) - 64/81\: L_1^3
  \nn \\ & & \mbox{} \quad
     + 992/81\: L_1^2 - 49.65\: L_1 + 11.32 + 51.94\: x 
     - x\,x_1 (44.52 + 11.05\: x)
     + 0.0647\:xL_0^4
  \nn \\ & & \mbox{} \quad
     - L_0 L_1\, ( 39.99 + 5.103\: L_0 - 16.30\: L_1) 
     - 16.00\: L_0 - 2848/243\: L_0^2 - 368/243\: L_0^3 \, \big\}
  \nn \\ &+& \mbox{} \!\! fl_{02}\: \nf \: \big\{
     2.147\: L_1^2 - 24.57\: L_1 + 48.79 - x_1 (242.4 - 150.7\: x) 
     - L_0 L_1\, (81.70 + 9.412\: L_1) 
  \nn \\ & & \mbox{} \quad
     + xL_0 \, (218.1 + 82.27\,L_0^2)
     - 477.0\: L_0 - 113.4\: L_0^2 + 17.26\: L_0^3 - 16/27\: L_0^5
     \,\big\} \: x_1 \:\: .
\end{eqnarray}
\vspace*{-10mm}
\end{table*}

Likewise the (experimentally irrelevant) coefficient functions for DIS via
Higgs exchange have been computed to three loops in the heavy top-quark limit,
see~Table~1. The $x\ra 1$ limit of these (also still unpublished) expressions 
forms the basis for the $H\;\!\!gg$ results in the next two sections.

\section{Massless quark and gluon form factors}

The form factors of quarks and gluons are gauge invariant (but infrared
divergent) parts of the perturbative corrections to inclusive hard scattering
processes. They summarize the QCD corrections to the $qqX$ and $ggX$ vertices
with a colour-neutral particle $X$ of either space-like or time-like momentum
$q$. These quantities are also key ingredients in the infrared factorization
of general higher-order amplitudes~\cite{Catani:1998bh,Sterman:2002qn}.

\vspace{0.5mm}
The relevant amplitude for the space-like $\gamma^{\ast}\!qq$ case, directly
entering the photon-exchange coefficient functions discussed above, reads
\beq
\label{eq:fmudeq}
\Gamma_\mu  \: = \: {\rm i} e_{\rm q}\,
\bigl({\bar u}\, \gamma_{\mu\,} u\bigr)\, {\cal F}_{\rm q} (\as,Q^2)
\eeq
where $e_{\rm q}$ represents the quark charge and $Q^2 = -q^2$ the virtuality
of the photon.  The gauge-invariant scalar function ${\cal F}_{\rm q}$ is the
space-like quark form factor which can be calculated order by order in the
strong coupling in dimensional regularization with $D=4-2\ep$.

The corresponding $H\;\!\!gg$ vertex defining ${\cal F}_{\:\!\!\rm g}$ is an 
effective interaction in the heavy top-quark limit,
\beq
\label{eq:LggH}
   {\cal L}_{\rm eff} \; = \; - \frac{1}{4} \, C_H \, H \,
   G^{\,a}_{\mu\nu} G^{\,a,\mu\nu} \:\: ,
\eeq
where $G^{\,a}_{\mu\nu}$ denotes the gluon field strength tensor, and the
coefficient $C_H$ includes all QCD corrections to the top-quark loop, known to 
N$^3$LO~\cite{Chetyrkin:1997un}, see also Refs.~\cite
{Schroder:2005hy,Chetyrkin:2005ia}.

\vspace{0.5mm}
The quark and gluon form factors were directly calculated at two loops in 
Refs.~\cite{Matsuura:1987wt} and \cite{Harlander:2000xx} to order $\ep^0$, 
respectively, and extended to (all) higher powers of $\ep$ in Refs.~\cite
{Moch:2005id,Moch:2005tm,Gehrmann:2005pd}. For the status of a direct 
three-loop calculation see Ref.~\cite{Gehrmann:2006wg}.

\vspace{0.5mm}
The $\ep^{-6} \,\ldots\, \ep^{-1}$ pole terms of the three-loop form factors 
can be extracted from the third-order coefficient functions for DIS
\cite{Moch:2005id,Moch:2005tm}. For this purpose we consider the bare 
(unrenormalized and unfactorized) partonic structure functions $F^{\rm b}$ for
$\,\gamma^{\ast} {\rm q} \ra {\rm q}X\,$ and $\,\phi^{\,\ast} {\rm g} \ra
{\rm g}X\,$ in the limit $x \ra 1$. Keeping, at each order $\as^{\,n}$, only 
the singular pieces proportional to $\delta(1-x)$ and the +-distributions
\beq
\label{eq:DDdef}
  \DD_{\:\!l} \: = \: \left[ \frac{\ln^{\,l}(1-x)}{(1-x)} \right]_+ \:\: ,
  \quad l \: = \: 1,\,\ldots\, 2n-1 \; , \;\;
\eeq
these results are compared to the general structure of the $n$-th order 
contribution $F^{\rm b}_n$ in terms of the $l$-loop form factors ${\cal F}_l$ 
and the corresponding real-emission parts ${\cal S}_{l\,}$, 
\bea
F^{\rm b}_0
     &\!=\!& \delta(1-x) \nn \\[0.8mm]
F^{\rm b}_1
     &\!=\!& 2 {\cal F}_1\,\delta(1-x) + {\cal S}_1 \nn \\[0.8mm]
F^{\rm b}_2
     &\!=\!& \left(2 {\cal F}_2 + {\cal F}_1^{\,2} \right) \delta(1-x)
           + 2 {\cal F}_1 {\cal S}_1 + {\cal S}_2 \nn \\[0.8mm]
F^{\rm b}_3
     &\!=\!& \left(2 {\cal F}_3 + 2 {\cal F}_1 {\cal F}_2\right) \delta(1-x)
\nn \\[0.4mm] & & \mbox{\hspn}
           + \left(2 {\cal F}_2 + {\cal F}_1^{\,2} \right) {\cal S}_1
           + 2 {\cal F}_1 {\cal S}_2
           + {\cal S}_3\; . \quad
\label{eq:Fbdec}
\eea
The $x$-dependence of the real emission factors ${\cal S}_k$ in inclusive 
deep-inelastic scattering is of the form ${\cal S}_k(f_{k,\ep})$, where the 
$D$-dimensional +-dis\-tributions $f_{k,\ep}$ are defined by
\bea
\label{eq:Dplus}
  f_{k,\ep}(x) &\! =\! & \ep [\,(1-x)^{-1-k\ep}\,]_+
\nn \\
               &\! =\! & - {1 \over k}\, \delta(1-x) + \sum_{i=0}\,
                       {(-k \ep)^i \over i\, !}\,\ep\,\DD_{\:\!i}\; . \;\;
\eea
 
Thus, exploiting this particular analytical dependence, the $n$-loop form 
factor ${\cal F}_n$ can simply be extracted by the substitution
\beq
\label{eq:DD0subs}
  \DD_{\:\!0} \: \to \: {1 \over n \ep} \, \delta(1 - x)
  - \sum_{i=1}\, {(-n \ep)^i  \over i\, !}\, \DD_{\:\!i} \; ,
\eeq
once the combinations of lower-order quantities in Eq.~(\ref{eq:Fbdec}) 
-- determined before to a sufficiently high order in $\ep$ --
have been subtracted from the calculated results for $F^{\rm b}_n$. 
As $\delta(1-x)$ enters with a factor $1/\ep$, this extraction loses one power
in $\ep$. Hence from the third-order calculation to order $\ep^0$, as performed
for the coefficient functions, only the pole terms of ${\cal F}_3$ can be 
obtained in this manner. So far the $\ep^0$-term has only been derived for the
fermionic ($\nf$) part of the quark form factor \cite{Moch:2005tm}, via 
extending the corresponding DIS calculation \cite{Moch:2002sn} by one order in 
$\ep$.

\vspace*{0.5mm}
The reader is referred to Refs.~\cite{Moch:2005id,Moch:2005tm,Moch:2005dx} for
a further discussion of these results, including their very interesting 
structure in the context of the exponentiation of the form factors~\cite
{Collins:1980ih,Sen:1981sd,Magnea:1990zb,Ravindran:2004mb}. We would like to
note, however, that the $\as^3\, \ep^{-1}$ coefficients of the highest 
\mbox{$\zeta$-function} weights, $\z2\z3$ and $\z5$, agree with the results 
inferred, using the conjecture of Ref.~\cite{Kotikov:2004er}, from the 
calculation in ${\cal N}\!=\!4\,$ Super-Yang-Mills theory in Ref.~\cite
{Bern:2005iz}. 

\section{Higgs productions at (almost) {\boldmath N$^3$LO}}

Armed with the third-order splitting functions \cite{Moch:2004pa,Vogt:2004mw}
and the results of the previous section, one can now derive all soft-enhanced 
\mbox{(+-distribution)} contributions to the N$^3$LO cross sections 
$\hat{\sigma}^{(3)}$ for lepton-pair and Higgs boson production at 
colliders~\cite{Moch:2005ky,Laenen:2005uz}, see also Refs.~\cite
{Idilbi:2005ni,Ravindran:2005vv}. Analogous to Eq.~(\ref{eq:Fbdec}),
the soft limit of the bare cross sections $W^{\rm b}$ for 
$\, {\rm q}\bar{{\rm q}} \,\ra\, \gamma^{\ast} \,\ra\, l^+ l^-\,$ and
$\,{\rm gg} \,\ra\, H\,$ reads
\bea
\label{eq:Wbexp}
 W^{\rm b}_0
   &\!\!=\!\!& \delta(1-x) \nn \\[0.3mm]
 W^{\rm b}_1
   &\!\!=\!\!& 2\,\mbox{Re}\,{\cal F}_1\,\delta(1-x) + {\cal S}_1 \nn \\[0.3mm]
 W^{\rm b}_2
   &\!\!=\!\!& (2\,\mbox{Re}\,{\cal F}_2 
         + \left|{\cal F}_1\right|^2)\, \delta(1-x)
         + 2\,\mbox{Re}\, {\cal F}_1 {\cal S}_1 
\nn\\ & & \mbox{\hspn}
         + {\cal S}_2 \nn\\[0.3mm]
 W^{\rm b}_3
   &\!\!=\!\!& (2\,\mbox{Re}\,{\cal F}_3 
         + 2 \left|{\cal F}_1 {\cal F}_2 \right|) \, \delta(1-x)
\\
& & \mbox{\hspn}
         + (2\,\mbox{Re}\,{\cal F}_2 + \left|{\cal F}_1\right|^2) {\cal S}_1 
         + 2\,\mbox{Re}\,{\cal F}_1 {\cal S}_2 + {\cal S}_3\; , \nn
\eea
where, of course, ${\cal F}$ now denotes the time-like quark or gluon form 
factor, known by analytic continuation from $q^2 = -Q^2 < 0\,$ to $q^2 > 0$.

\vspace*{0.5mm}
The real-emission contributions ${\cal S}_k$ depend on the scaling variable 
$x = M_{\gamma^{\,\ast}\! ,\,H}^{\,2}/s$. Here the dependence of 
${\cal S}_k$ on $x$ is of the form ${\cal S}_k(f_{2k,\ep})$, i.e., 
\beq
\label{eq:SsoftDYH}
  {\cal S}_k \; = \; f_{2k,\ep}\, \sum_{l=-2k}^{\infty} 
  2k\, s_{k,l}^{}\: \ep^{\,l} \; .
\eeq
With the known time-like form factors, the expansion coefficients $s_{k,l}^{}$ 
of the soft function ${\cal S}_k$ can be derived recursively as far as they
are subject to the KLN cancellations and the mass-factorization structure
relating the remaining poles to the splitting functions (\ref{eq:P-exp}). 
Employing the results of Refs.~\cite{Moch:2005id,Moch:2005tm} and \cite
{Moch:2004pa,Vogt:2004mw}, the third-order terms $s_{3,-6}^{}\,\ldots\,s_{3,-1}$
can be obtained. Due to Eq.~(\ref{eq:Dplus}) this is sufficient to derive all
+-distribution contributions to the third-order coefficient functions. 
The explicit results can be found in Ref.~\cite{Moch:2005ky}.

\vspace*{0.5mm}
Focusing on Higgs production, we note that a good approximation (to about 10\% 
or less) to the double convolutions $g\ast g\ast \hat{\sigma}$ is obtained 
at NLO and NNLO~\cite{Harlander:2002wh,Anastasiou:2002yz,Ravindran:2003um}
by transforming to $N$-space and keeping only the $\ln^k N$ and $N^0$ terms 
arising from the \mbox{+-distributions} in $\hat{\sigma}^{(1)}$ and 
$\hat{\sigma}^{(2)}$.
Consequently the above results facilitate a sufficient approximation to the 
complete~N$^3$LO~correction, with a (conservative) error estimate of 20\%.

\begin{figure}[thb]
\vspace*{-8mm}
\centerline{\epsfig{file=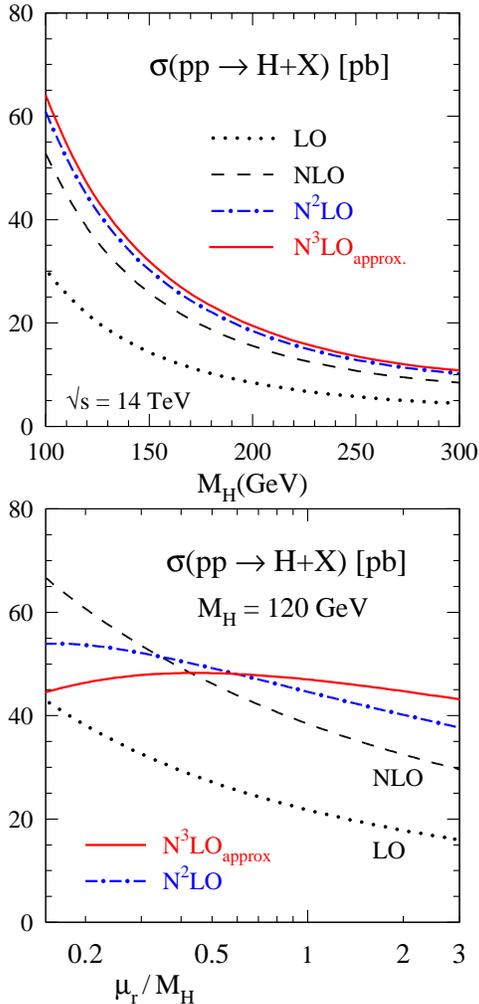,width=6.6cm,angle=0}\quad}
\vspace*{-9mm}
\caption{\label{mmvv-f4}
Perturbative expansion of the total cross section of Higgs production at
the LHC for the parton densities of Refs.~\cite{Martin:2001es,Martin:2002dr}.
Top:~dependence on the Higgs mass $M_{\rm H}$.
Bot\-tom: renor\-malization scale dependence for $M_{\rm H} =
120\mbox{ GeV}$.}
\vspace*{-1mm}
\end{figure}

\vspace{0.5mm}
The resulting predictions are illustrated for the LHC in Fig.\ref{mmvv-f4}, 
where all higher-order contributions have been calculated in the heavy 
top-quark approximation, but are normalized to the full lowest-order cross
sections. Considering these and other results \cite{Moch:2005ky}, 5\% at the 
LHC, and 7\% at the {\sc Tevatron}, appears to represent a conservative 
estimate of the improved cross-section uncertainty due to the truncation of 
the perturbation series at the (approximated) N$^3$LO.
 
\section{Summary}

We have computed the third-order coefficient functions for the most important
structure functions in deep-inelastic scattering 
\cite{Moch:2004xu,Vermaseren:2005qc,Vermaseren:2006zz}.
This first calculation of the three-loop corrections to one-scale partonic 
cross sections offers new insights into the process at hand, and facilitates 
improved determinations of the strong coupling constant $\as$. Moreover it has 
lead to further important third-order (and all-order resummed, cf.\ Ref.~\cite
{Moch:2005ba,Moch:2005dx}) results for the on-shell quark and gluon form 
factors \cite{Moch:2005id,Moch:2005tm} and the cross section for Higgs boson 
production at proton colliders \cite{Moch:2005ky}.
 
\section*{Acknowledgments}

The work of S.M. has been supported in part by the Helmholtz Gemeinschaft
under contract VH-NG-105. 
The work of J.V. has been part of the research program of the Dutch
Foundation for Fundamental Research of Matter (FOM).\\[3mm]

\end{document}